\documentclass[twocolumn, 
  aps, prl,
  amsmath,amssymb,
  longbibliography,
  ]{revtex4-1}
\usepackage{graphicx,color}
\usepackage{amsmath}
\usepackage{natbib}
\usepackage{epsfig}
\begin{document}

\title{\color{blue} Thermal conduction in two-dimensional complex plasma layers}

\author{Sergey A. Khrapak}
\email{Sergey.Khrapak@gmx.de}
\affiliation{Joint Institute for High Temperatures, Russian Academy of Sciences, 125412 Moscow, Russia; \\Institut f\"ur Materialphysik im Weltraum, Deutsches Zentrum f\"ur Luft- und Raumfahrt (DLR), 82234 We{\ss}ling, Germany
}

\begin{abstract}
A simple vibrational model of heat transfer in two-dimensional (2D) fluids relates the heat conductivity coefficient to the longitudinal and transverse sound velocities, specific heat, and the mean interatomic separation. This model is demonstrated not to contradict the available experimental and numerical data on heat transfer in 2D complex plasma layers. Additionally, the heat conductivity coefficient of a 2D one-component plasma with a logarithmic interaction is evaluated.      
\end{abstract}

\date{\today}

\maketitle

Recently, a vibrational model of heat transfer in simple three-dimensional (3D) liquids with soft interparticle interactions has been discussed~\cite{HT_OCP}. The model has been applied to quantify heat transfer in a strongly coupled one-component plasma (OCP) and Lennard-Jones fluids and remarkable agreement with available numerical results has been reported. If there is some truth in this simple model, it can be very straightforwardly generalized to two-dimensional (2D) systems. The purpose of this Letter is to present such a generalization and to compare its prediction with existing results from experiments and simulations on heat transfer in 2D complex plasma layers. Additionally, we make a prediction about the yet unknown heat conduction coefficient of a 2D OCP.

The problem of transport coefficients in 2D systems has been and remains a rather controversial issue. The absence of valid transport coefficients in 2D systems was predicted based on the analysis of the velocity autocorrelation function and of the kinetic
parts of the correlation functions for the shear viscosity and the heat conductivity~\cite{ErnstPRL1970}. Molecular-dynamics simulations of the 2D classical electron system (2D Coulomb fluid) yielded indications for the existence of a self-diffusion coefficient~\cite{HansenPRL1979}. Strong indications of normal self-diffusion in 2D Yukawa fluids at sufficiently long time scales were also reported~\cite{OttPRL2009}. Existence of finite shear viscosity coefficients of strongly coupled 2D Yukawa fluids was demonstrated in experiments with complex (dusty) plasma monolayers and molecular dynamics (MD) simulations~\cite{DonkoMPLB2007}. In a dedicated study of Ref.~\cite{DonkoPRE2009} it was observed that the  self-diffusion coefficient exists at sufficiently high temperatures, the viscosity coefficient exists at sufficiently low temperatures, but not in the opposite limits (so that the Stokes-Einstein relation does not hold in 2D). The thermal conductivity coefficient does not appear to exist at high temperature. For low temperatures no definite conclusion could be drawn, because of the technical challenges posed by signal-to-noise ratios and a long initial decay of the corresponding correlation function~\cite{DonkoPRE2009}.

\begin{figure}
\includegraphics[width=8.cm]{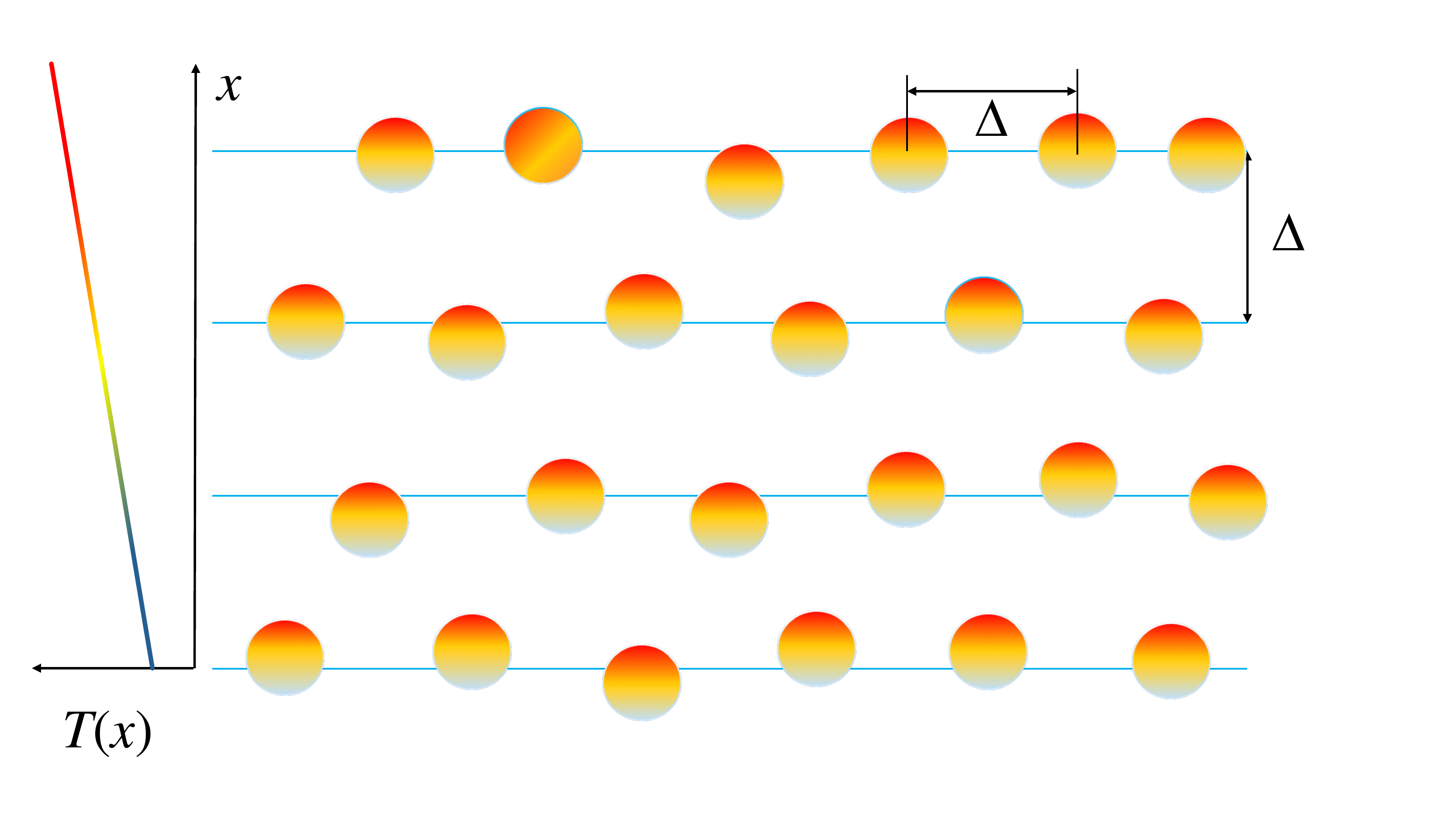}
\caption{(Color online) Sketch of a two-dimensional fluid-like structure. The average inter-particle separation is $\Delta = n^{-1/2}$. The temperature increases from bottom to top.}
\label{Fig1}
\end{figure}

An outstanding review of thermal conduction in classical low-dimensional lattices can be found in Ref.~\cite{LepriPR2003}. The mode-coupling theory predicts finite thermal conductivity coefficient in 3D and its divergence in the thermodynamic limit for lower dimensions: $\lambda\propto \ln N$ (in 2D) and $\lambda\propto N^{2/5}$ (in 1D), where $N$ is the number of particles in the system. Nevertheless, numerous indications in favor of the finiteness of $\lambda$ in the thermodynamic limit of low dimensional systems have also been reported over the years~\cite{LepriPR2003}.

On the experimental side, finite values of the thermal conductivity coefficient were reported for single layers of crystalline, melted, and fluid complex (dusty) plasmas as measured in dedicated experiments~\cite{NunomuraPRL2005,FortovPRE2007,NosenkoPRL2008,
GoreePPCF2013,DuPRE2014}.
Numerical simulations of heat conduction in 2D Yukawa systems (which serve as a first approximation for real complex plasmas) were also performed for conditions representative to those in laboratory experiments~\cite{HouJPA2009,KhrustalyovPRE2012,ShahzadPoP2015}. So how do these results compare with theoretical expectations?
         
First, we generalize a simple vibrational model of heat transfer to the 2D situation. 2D fluid is approximated by a string-like structure with strings perpendicular to the temperature gradient and separated by the distance $\Delta = n^{-1/2}$. The average particle separation in each string is also $\Delta$. A sketch of the considered structure is shown in Fig.~\ref{Fig1}. It resembles that of a three-dimensional layered structure, but here the particles tend to form a disordered quasi-triangular lattice. When a temperature gradient is applied, the average difference in energy between the atoms of adjacent strings is $\Delta (dU/dx)$, where $U$ is the internal energy per particle. The energy between successive strings is transferred when two vibrating particles from adjacent strings ``collide'' (that is they approach by a distance that is shorter than the average interparticle separation). The characteristic energy transfer frequency is given by the average vibrational frequency of a particle, $\langle \omega \rangle/2\pi$.
The energy flux per unit string length is
$dQ/dt=-(\langle \omega\rangle/2\pi)(dU/dx)$,
where the minus sign indicates that the heat flow is down the temperature gradient. Comparing this with Fourier's law for the heat $dQ/dt=-\lambda(dT/dx)$,
where $\lambda$ is the thermal conductivity coefficient (scalar in isotropic liquids), we get
\begin{equation}\label{Cond1}
\lambda=c_V\frac{\langle \omega \rangle}{2\pi},
\end{equation}
where $c_V=dU/dT$ is the reduced heat capacity at constant volume. The temperature is measured in energy units ($k_{\rm B}=1$) and hence the thermal conductivity coefficient is expressed in units of frequency.

Equation (\ref{Cond1}) represents a general formula of the thermal conductivity coefficient in 2D fluids, which can be further simplified using special assumptions about the vibrational spectrum. In the simplest Einstein approximation all particles vibrate with the same (Einstein) frequency $\Omega_{\rm E}$ (on time scales shorter than structural rearrangement time scale) and, hence, $\langle \omega \rangle = \Omega_{\rm E}$. We obtain 
\begin{equation}\label{CondE}
\lambda=c_V\frac{\Omega_{\rm E}}{2\pi},
\end{equation}
which can be considered as a 2D analogue of the 
Horrocks and McLaughlin formula~\cite{Horrocks1960}. Moreover, in the vicinity of the fluid-solid phase transition we have $c_V\simeq 2$ to a good approximation and hence
\begin{equation}\label{CondE1}
 \lambda\simeq \frac{\Omega_{\rm E}}{\pi}.
\end{equation} 
This can be used as a rough estimate of the thermal conductivity coefficient of 2D melts.  

%As an alternative we can use an acoustic spectrum, $\omega=c_{\rm s}k$, supplemented by an appropriate cut-off of the wave numbers $k_{\rm max}$. Then, the standard averaging procedure results an analogue of the Bridgman equation (\ref{Bridgman}). Thus, the two expressions from the Introduction can be considered as two special simplifications of the more general expression (\ref{Cond1}).   

%As a further improvement, let us consider a Debye spectrum, characterized by the vibrational density of states that is proportional to $\omega$ in 2D, $g(\omega)\propto \omega$. We get
%\begin{equation}
%\langle \omega \rangle=\int_0^{\omega_{\rm D}}g(\omega)\omega d\omega \left[\int_0^{\omega_{\rm D}}g(\omega)d\omega\right]^{-1}=\frac{2}{3}\omega_{\rm D},
%\end{equation}
%where $\omega_{\rm D}$ is the cutoff Debye frequency. 

%The latter can be estimated from the condition 
%\begin{displaymath}
%\Omega_{\rm E}^2=\langle \omega^2 \rangle = \frac{3}{5}\omega_{\rm D}^2,
%\end{displaymath}
%which yields $\langle \omega \rangle= 0.968 \Omega_{\rm E}$, which is again close to the result by Horrocks and McLaughlin. 
  
As a more accurate approximation we consider conventional long-wavelength spectrum containing one longitudinal and one transverse acoustic-like modes with
\begin{equation}\label{dispersion}
\omega_l^2\simeq c_{l}^2k^2; \quad \omega_t^2\simeq c_{t}^2k^2,
\end{equation} 
where $c_l$ and $c_t$ are the longitudinal and transverse sound velocities. The standard averaging procedure yields
\begin{equation}\label{integral}
\langle \omega \rangle = \frac{1}{4\pi n}\int_0^{k_{\rm max}}k dk\left[\omega_l(k)+\omega_t(k)\right]=\frac{k_{\rm max}^3(c_l+c_t)}{12\pi n}.
\end{equation} 
The cutoff wavenumber $k_{\rm max}$ is chosen to provide $n$ oscillations for each collective mode, that is $\pi(k_{\rm max}/2\pi)^2=n$ (to ensure that we have 2$N$ normal modes for $N$ particles in 2D). The average vibrational frequency becomes 
\begin{equation}\label{freq}
\langle \omega \rangle = \frac{2\sqrt{\pi}}{3}\frac{c_l+c_t}{\Delta}\simeq 1.18 \frac{c_l+c_t}{\Delta}.
\end{equation}
This leads to the following simple approximation for the thermal conductivity coefficient of simple 2D fluids
\begin{equation}\label{CondC}
\lambda\simeq 0.1878 c_V \frac{c_l+c_t}{\Delta}.
\end{equation} 
If we treat the sum $c_l+c_t$ as some effective sound velocity $c_{s}$, then Eq.~(\ref{CondC}) can be considered as a 2D analogue of the Bridgman's equation for the thermal conductivity of liquids~\cite{Bridgman1923,BirdBook}.

In arriving at (\ref{freq}) two additional simplifications have been employed. First, the kinetic terms were not included in acoustic dispersion relations of Eq.~(\ref{dispersion}). These are numerically small at fluid densities and can be omitted. Second, the existence of a $k$-gap in the dispersion relation of the transverse collective mode is not taken into account. This ``$k$-gap'' implies a minimum (critical) wave number, $k_*$, below which transverse (shear) waves cannot propagate. It occurs in fluids both in 3D and 2D. The critical wave number tends to decrease with increase of density and decrease of temperature (on approaching the liquid-solid phase transition)~\cite{MurilloPRL2000,GoreePRE2012,YangPRL2017,KhrapakJCP2018,
KhrapakJCP2019,KryuchkovSciRep2019}.
Since the contribution from the small $k$-region is not essential in the present context, the existence of the $k$-gap does not affect the obtained results explicitly (as long as $k_*\ll k_{\rm max}$, of course). At the same time, the $k$-gap itself is directly related to the heat capacity $c_V$ and thus it affects the magnitude of the thermal conductivity coefficient~\cite{KryuchkovPRL2020}.     

It is now possible to compare the simple expressions (\ref{CondE}) and (\ref{CondC}) with the available results from experiments and simulations. We start with the experiment performed by Nosenko~\cite{NosenkoPRL2008}. In this experiment the heat transport in a 2D complex plasma undergoing a solid-fluid phase transition was
studied. A single layer of highly charged polymer microspheres was suspended in a plasma
sheath. A part of this lattice was heated by two counterpropagating focused laser beams that moved
rapidly around in the lattice and provided short intense random kicks to the particles. Above a threshold,
the lattice locally melted. The spatial profiles of the particle kinetic temperature were analyzed to find a
thermal conductivity coefficient. For the parameters investigated the numerical value of $\lambda\simeq 21$ s$^{-1}$ was obtained~\cite{NosenkoPRL2008,NosenkoPC}. All the parameters necessary for comparison with the present model were measured experimentally. The longitudinal and shear waves had the sound velocities $c_l\simeq 28.7$ mm/s and $c_t\simeq 5.4$ mm/s. The interparticle separation $\Delta$, measured from the radial distribution function $g(r)$ was about $0.7$ mm in the center of the particle layer. Substituting these numbers into Eq.~(\ref{CondC}) yields $\lambda\simeq 18.3$ s$^{-1}$ if $c_V\simeq 2$ is assumed. Actually, the specific heat $c_V$ is known to somewhat exceed the harmonic value 2 (in 2D) near the fluid-solid phase transition due to anharmonic effects, both in the fluid and solid phases~\cite{KhrapakCPP2016}. This would shift the theoretical value even closer to the experimental result.     

Next, let us consider some of the available results from numerical simulations. All simulations correspond to 2D Yukawa systems, as the first approximation of real dusty plasma layers, and some background information is necessary. 2D Yukawa systems represent point-like particles interacting via the pairwise interaction potential $\phi(r)=(Q^2/r)\exp(-r/\lambda_{\rm s})$, where $Q$ is the particle charge and $\lambda_{\rm s}$ is the screening length. Equilibrium structural properties of these systems are fully characterized by the two dimensionless parameters: the coupling parameter $\Gamma=Q^2/aT$ and the screening parameter $\kappa=a/\lambda_{\rm s}$, where $a=(\pi n)^{-1/2}$ is the 2D Wigner-Seitz radius,  $T$ is the temperature in energy units ($k_{\rm B}=1$), and $n$ is the (areal) density. Equilibrium thermodynamic properties and phase diagram of 2D Yukawa systems are relatively well investigated (see e.g. Refs.~\cite{HartmannPRE2005,KryuchkovJCP2017,KhrapakPRR2020}). Dynamics is also affected by the damping coefficient $\gamma$, which is mainly related to the presence of neutral atoms in weakly ionized dusty plasma (and is often referred to as Epstein drag).  

Hou and Piel (HP)~\cite{HouJPA2009} conducted non-equilibrium Brownian dynamics simulations to study heat transfer in strongly coupled 2D Yukawa systems, for parameters close to those in real experiments~\cite{NunomuraPRL2005,NosenkoPRL2008}. In HP simulation the system was first brought to an equilibrium with a desired temperature in either liquid or solid state. Then half of the system was heated to a higher temperature by applying a Gaussian white noise with desired strength. The evolution of the temperature
profile and the heat flux were recorded. A steady state was approached after a substantially long period. Fitting
the stationary temperature profile to a simple analytical model yielded the heat conductivity coefficient. In their simulations the screening parameter was kept constant at $\kappa = 1$, while $\Gamma$ and $\gamma$ were varied to examine different equilibrium states and different
damping rates. HP reported $\lambda\simeq 0.35 \omega_{\rm p}$ for the parameters of the experiment in Ref.~\cite{NosenkoPRL2008}, where $\omega_{\rm p}=\sqrt{2\pi Q^2n/ma}$ is the 2D plasma frequency (it was observed that $\lambda$ rises slightly with the decrease of $\gamma$ and tends to  $\lambda\simeq 0.4 \omega_{p}$ in the frictionless limit). From the theoretical side, we have $\Omega_{\rm E}\simeq 0.5 \omega_{\rm p}$ at $\kappa = 1$ (see Fig. 4 from Ref.~\cite{KhrapakPoP2018}). Substituting this into Eq.~(\ref{CondE1}) we get $\lambda=0.16\omega_{\rm p}$. Alternatively, we can substitute the sound velocities $c_l\simeq 0.81\omega_{\rm p}a$ and $c_t\simeq 0.23 \omega_{\rm p}a$ of strongly coupled 2D Yukawa fluids at $\kappa=1$~\cite{DonkoJPCM2008,KhrapakPoP2016_Relations} to obtain $\lambda\simeq 0.22\omega_{\rm p}$. Both theoretical approximations are somewhat lower than HP numerical result.    

Khrustalyov and Vaulina (KV)~\cite{KhrustalyovPRE2012} 
studied thermal conductivity of equilibrium 2D Yukawa systems by means of Langevin molecular
dynamics simulations. They calculated the thermal conductivity coefficient from the
Green-Kubo expression. The influence of frictional dissipation on the heat transfer was investigated. They expressed the thermal conductivity coefficient (unconventionally) in units of $n\omega^* r_p^2$, where $\omega^*$ is some characteristic frequency of collisions between the charged macroparticles and $r_p$ is some characteristic interparticle separation. In the frictionless limit they obtained $\lambda/n\omega^*r_p^2\simeq 0.86$ near the fluid-solid phase transition, independently of $\kappa$ (for $\kappa\leq 4$). From the information presented in Ref.~\cite{KhrustalyovPRE2012} we can estimate $r_p/a\simeq 1.4$ and $\omega^*\simeq 0.34 \omega_{\rm p}$ in the weakly screened limit ($\kappa\lesssim 1$)~\cite{Gamma}. This yields $\lambda\simeq 0.18\omega_{\rm p}$, which is compatible with the theoretical approximation presented above. For example, in the weakly screening limit $\kappa\ll 1$ we have $\Omega_{\rm E}\simeq 0.6 \omega_{\rm p}$ (see Fig. 4 from Ref.~\cite{KhrapakPoP2018}). Using Eq.~(\ref{CondE1}) we immediately get $\lambda=0.19\omega_{\rm p}$. 

A homogenous nonequilibrium molecular dynamics simulation was used to compute the thermal conductivity coefficient  of 2D strongly coupled Yukawa fluids by Shahzad and He (SH)~\cite{ShahzadPoP2015}. They used two different normalizations for $\lambda$ (by employing plasma and Einstein frequencies, respectively) and attempted to compare their results with those from earlier simulations (including 3D results) and experiments. The comparison is not conclusive: The results for $\lambda$ from HP simulation appear smaller than those from KV simulation, contrary to the actual state of affairs. Nevertheless, the results presented in Fig.~5 of Ref.~\cite{ShahzadPoP2015} seem to indicate that the quantity $\lambda/n\Omega_{\rm E}a^2$ approaches unity as the crystallization point is approached. This finding is perfectly consistent with Eq.~(\ref{CondE1}) since $\pi na^2=1$. 

Thus, the vibrational model of heat conductivity in 2D complex plasma layers is not inconsistent with the results from experiments and simulations for which comparison is possible. Except one (HP simulation) case the model is able to reproduce closely the numerical values of the heat conductivity coefficient. 
In the HP case~\cite{HouJPA2009} the theoretical result is more than $30\%$ lower than that from simulation and the reason for this is unclear.   

What becomes also clear after careful analysis of the available simulations is that there is an obvious deficiency of the numerical data (compared to the 3D case~\cite{SalinPoP2003,SalinPRL2002,DonkoPRE2004,
FaussurierPRE2003,FaussurierPRE2004,ShahzadPoP2012,
ShahzadCPP2012,ScheinerPRE2019}). Only several state points have been investigated so far. No careful analysis of general trends and tendencies across the phase diagram has been performed. For instance, even the temperature dependence of $\lambda$ in the strongly coupled regime remains unclear (apart from the fact that it is rather weak). Conflicting predictions have been made: In Refs.~\cite{KhrustalyovPRE2012} and \cite{ShahzadPoP2015} the data comply with a monotonous decrease of $\lambda$ on approaching the freezing point; In Ref.~\cite{HouJPA2009} slight increase of $\lambda$ was suggested. The vibrational model predicts weak increase of $\lambda$ on approaching the fluid-solid transition, mainly due to increase of the specific heat $c_V$. 

The dependence of $\lambda$ on the system size was not investigated systematically, probably because the interest was mainly concentrated on conditions close to those in laboratory experiments. In experiments~\cite{NunomuraPRL2005,NosenkoPRL2008} 5000 - 10000 particles formed a 2D horizontal layer, of which positions and movements of about 900 were traced using a video camera. HP used 10000 particles in their simulation, KV varied the number of particles from 256 to 1225, SH used  $N=1024$, 4096, 14 400, and 22 500 particles, but did not provide any convincing conclusions regarding the system size dependence of $\lambda$. It cannot be completely excluded that some difference between the reported results may arise from the system size effects.                
  
Overall, relatively little is still known about thermal conductivity in 2D Yukawa systems (and about 2D systems in general). Although the vibrational model does not contradict the results reported so far, more results are needed to get more confidence regarding its applicability conditions. Experiments in large 2D complex plasma systems~\cite{NosenkoAIP2018} as well as new simulations covering
different regions of the phase space and addressing the effect of system size are required to fulfil this program.    

To provide one more test case for potential future comparison, we will now calculate a yet unknown coefficient of thermal conductivity of a 2D one-component plasma (OCP) with a logarithmic interaction potential. The logarithmic potential, $\phi(r)=-Q^2\ln(r/a)$, emerges from the solution of the 2D Poisson equation and represents the interaction between infinite charged filaments ($Q^2$ is expressed in energy units). The point charges are immersed into a rigid neutralizing background to guarantee system stability and finite thermodynamical quantities. This system should not be confused with a system of point charges interacting via the conventional 3D  Coulomb potential ($\propto 1/r$), whose motion is restricted to a 2D surface (which is also often referred to as OCP and represents the $\kappa = 0$ limit of Yukawa systems considered above).

% The system is characterized by ultra-soft interactions between the particles and is of interest from the fundamental point of view as an opposite limit of the celebrated hard sphere (hard disc in 2D) model. 

\begin{figure}
\includegraphics[width=7.5cm]{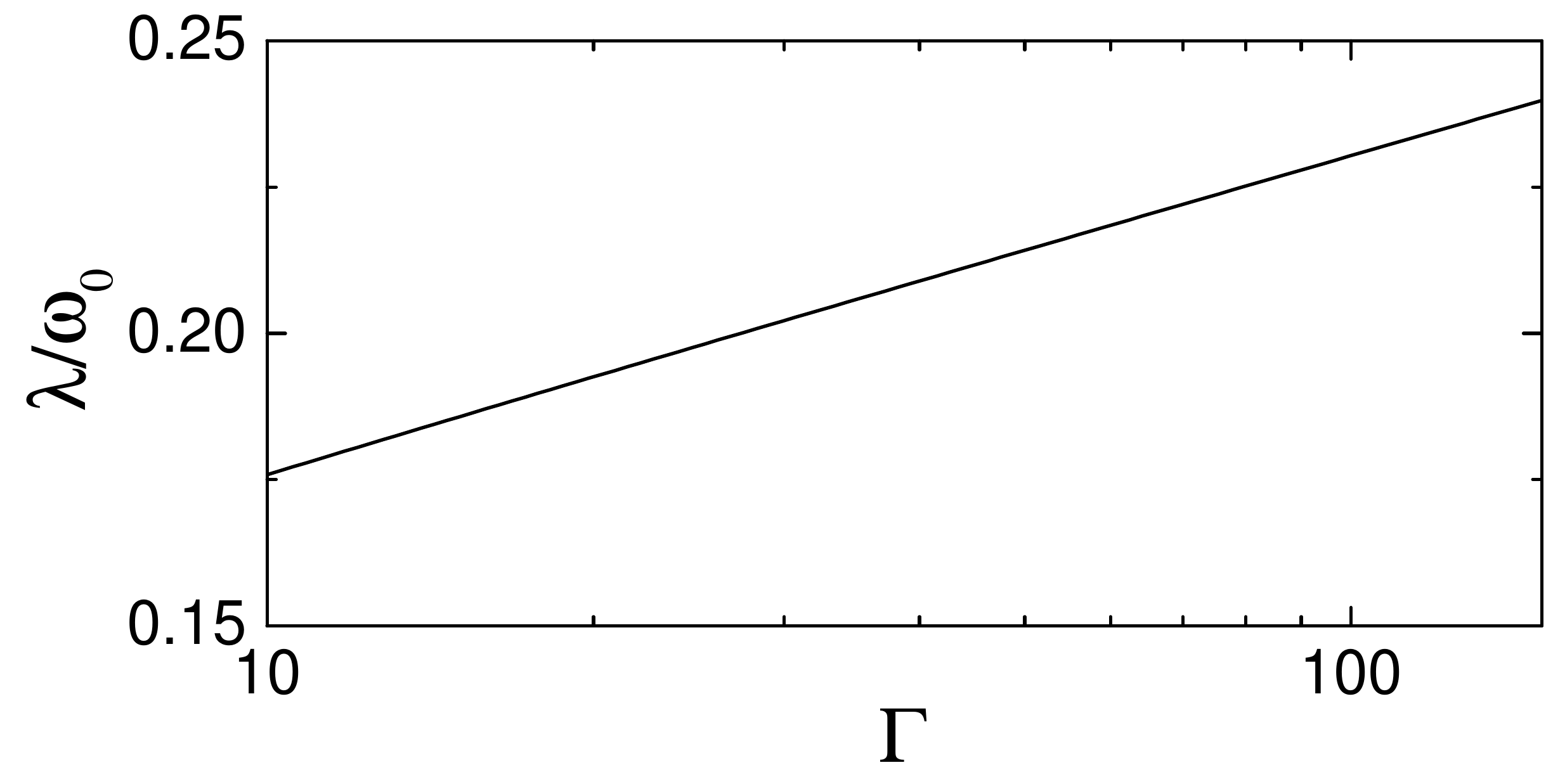}
\caption{Reduced thermal conductivity coefficient of a 2D one-component plasma with logarithmic interactions, $\lambda/\omega_0$, as a function of the coupling parameter $\Gamma$.}
\label{Fig2}
\end{figure}

Thermodynamic properties of the 2D OCP are governed by the coupling parameter $\Gamma=Q^2/T$; dynamical properties are expressed using the 2D frequency scale $\omega_0=\sqrt{2\pi Q^2 n/m}$~\cite{Alastuey1981,Caillol1982,Leeuw1982,Leeuw1983}. 
The dispersion relations of the longitudinal and transverse modes (long-wavelength branch) at strong coupling can be to a good accuracy approximated by~\cite{KhrapakPoP2016_2DOCP}
\begin{equation}
\frac{\omega_{l,t}^2}{\omega_0^2}=\frac{1}{2}\pm\frac{J_1(q)}{q},
\end{equation}    
where $q=ka$ is the normalized wave-number, $J_1(x)$ is the Bessel function of the first kind, and plus (minus) sign corresponds to the longitudinal (transverse) mode, respectively. Averaging according to Eq.~(\ref{integral}) yields $\langle \omega \rangle \simeq 0.629\omega_0$. 
The specific heat can be calculated from the thermal component of the internal energy, for which we use an approximation of Eq.~(20) from Ref.~\cite{KhrapakCPP2016}. The resulting dependence of the reduced thermal conductivity coefficient on $\Gamma$ in the strongly coupled regime is shown in Fig.~\ref{Fig2}. An order of magnitude estimate can be done with the help of Eq.~(\ref{CondE1}) taking into account that $\Omega_{\rm E}=\omega_0/\sqrt{2}$ for the 2D OCP, which results in $\lambda\simeq 0.23\omega_0$ near freezing. The obtained result can be tested when numerical results on the thermal conductivity of 2D OCP become available.

To conclude, a vibrational model of heat conductivity in 2D fluids has been presented and discussed. It does not contradict the available results from experiments and numerical simulations of 2D complex plasma layers. Nevertheless, more detailed comparison is clearly warranted and some suggestions for future research directions have been made.        

I would like to thank Volodimir Nosenko for communications during the preparation of this manuscript and its careful reading.

\bibliography{TC_Ref}

\providecommand{\noopsort}[1]{}\providecommand{\singleletter}[1]{#1}%
\begin{thebibliography}{48}%
\makeatletter
\providecommand \@ifxundefined [1]{%
 \@ifx{#1\undefined}
}%
\providecommand \@ifnum [1]{%
 \ifnum #1\expandafter \@firstoftwo
 \else \expandafter \@secondoftwo
 \fi
}%
\providecommand \@ifx [1]{%
 \ifx #1\expandafter \@firstoftwo
 \else \expandafter \@secondoftwo
 \fi
}%
\providecommand \natexlab [1]{#1}%
\providecommand \enquote  [1]{``#1''}%
\providecommand \bibnamefont  [1]{#1}%
\providecommand \bibfnamefont [1]{#1}%
\providecommand \citenamefont [1]{#1}%
\providecommand \href@noop [0]{\@secondoftwo}%
\providecommand \href [0]{\begingroup \@sanitize@url \@href}%
\providecommand \@href[1]{\@@startlink{#1}\@@href}%
\providecommand \@@href[1]{\endgroup#1\@@endlink}%
\providecommand \@sanitize@url [0]{\catcode `\\12\catcode `\$12\catcode
  `\&12\catcode `\#12\catcode `\^12\catcode `\_12\catcode `\%12\relax}%
\providecommand \@@startlink[1]{}%
\providecommand \@@endlink[0]{}%
\providecommand \url  [0]{\begingroup\@sanitize@url \@url }%
\providecommand \@url [1]{\endgroup\@href {#1}{\urlprefix }}%
\providecommand \urlprefix  [0]{URL }%
\providecommand \Eprint [0]{\href }%
\providecommand \doibase [0]{http://dx.doi.org/}%
\providecommand \selectlanguage [0]{\@gobble}%
\providecommand \bibinfo  [0]{\@secondoftwo}%
\providecommand \bibfield  [0]{\@secondoftwo}%
\providecommand \translation [1]{[#1]}%
\providecommand \BibitemOpen [0]{}%
\providecommand \bibitemStop [0]{}%
\providecommand \bibitemNoStop [0]{.\EOS\space}%
\providecommand \EOS [0]{\spacefactor3000\relax}%
\providecommand \BibitemShut  [1]{\csname bibitem#1\endcsname}%
\let\auto@bib@innerbib\@empty
%</preamble>
\bibitem [{\citenamefont {Khrapak}(2021)}]{HT_OCP}%
  \BibitemOpen
  \bibfield  {author} {\bibinfo {author} {\bibfnamefont {S.}~\bibnamefont
  {Khrapak}},\ }\bibfield  {title} {\enquote {\bibinfo {title} {Vibrational
  model of thermal conduction for fluids with soft interactions},}\ }\href
  {\doibase 10.1103/PhysRevE.103.013207} {\bibfield  {journal} {\bibinfo
  {journal} {Phys. Rev. E}\ }\textbf {\bibinfo {volume} {103}},\ \bibinfo
  {pages} {013207} (\bibinfo {year} {2021})}\BibitemShut {NoStop}%
\bibitem [{\citenamefont {Ernst}\ \emph {et~al.}(1970)\citenamefont {Ernst},
  \citenamefont {Hauge},\ and\ \citenamefont {van Leeuwen}}]{ErnstPRL1970}%
  \BibitemOpen
  \bibfield  {author} {\bibinfo {author} {\bibfnamefont {M.~H.}\ \bibnamefont
  {Ernst}}, \bibinfo {author} {\bibfnamefont {E.~H.}\ \bibnamefont {Hauge}}, \
  and\ \bibinfo {author} {\bibfnamefont {J.~M.~J.}\ \bibnamefont {van
  Leeuwen}},\ }\bibfield  {title} {\enquote {\bibinfo {title} {Asymptotic time
  behavior of correlation functions},}\ }\href {\doibase
  10.1103/physrevlett.25.1254} {\bibfield  {journal} {\bibinfo  {journal}
  {Phys. Rev. Lett.}\ }\textbf {\bibinfo {volume} {25}},\ \bibinfo {pages}
  {1254--1256} (\bibinfo {year} {1970})}\BibitemShut {NoStop}%
\bibitem [{\citenamefont {Hansen}\ \emph {et~al.}(1979)\citenamefont {Hansen},
  \citenamefont {Levesque},\ and\ \citenamefont {Weis}}]{HansenPRL1979}%
  \BibitemOpen
  \bibfield  {author} {\bibinfo {author} {\bibfnamefont {J.~P.}\ \bibnamefont
  {Hansen}}, \bibinfo {author} {\bibfnamefont {D.}~\bibnamefont {Levesque}}, \
  and\ \bibinfo {author} {\bibfnamefont {J.~J.}\ \bibnamefont {Weis}},\
  }\bibfield  {title} {\enquote {\bibinfo {title} {Self-diffusion in the
  two-dimensional, classical electron gas},}\ }\href {\doibase
  10.1103/physrevlett.43.979} {\bibfield  {journal} {\bibinfo  {journal} {Phys.
  Rev. Lett.}\ }\textbf {\bibinfo {volume} {43}},\ \bibinfo {pages} {979--982}
  (\bibinfo {year} {1979})}\BibitemShut {NoStop}%
\bibitem [{\citenamefont {Ott}\ and\ \citenamefont
  {Bonitz}(2009)}]{OttPRL2009}%
  \BibitemOpen
  \bibfield  {author} {\bibinfo {author} {\bibfnamefont {T.}~\bibnamefont
  {Ott}}\ and\ \bibinfo {author} {\bibfnamefont {M.}~\bibnamefont {Bonitz}},\
  }\bibfield  {title} {\enquote {\bibinfo {title} {Is diffusion anomalous in
  two-dimensional yukawa liquids?}}\ }\href {\doibase
  10.1103/physrevlett.103.195001} {\bibfield  {journal} {\bibinfo  {journal}
  {Phys. Rev. Lett.}\ }\textbf {\bibinfo {volume} {103}},\ \bibinfo {pages}
  {195001} (\bibinfo {year} {2009})}\BibitemShut {NoStop}%
\bibitem [{\citenamefont {Donko}\ \emph {et~al.}(2007)\citenamefont {Donko},
  \citenamefont {Hartmann},\ and\ \citenamefont {Goree}}]{DonkoMPLB2007}%
  \BibitemOpen
  \bibfield  {author} {\bibinfo {author} {\bibfnamefont {Z.}~\bibnamefont
  {Donko}}, \bibinfo {author} {\bibfnamefont {P.}~\bibnamefont {Hartmann}}, \
  and\ \bibinfo {author} {\bibfnamefont {J.}~\bibnamefont {Goree}},\ }\bibfield
   {title} {\enquote {\bibinfo {title} {Shear viscosity of strongly-coupled
  two-dimensional yukawa liquids:. experiment and modeling},}\ }\href {\doibase
  10.1142/s0217984907013948} {\bibfield  {journal} {\bibinfo  {journal} {Mod.
  Phys. Lett. B}\ }\textbf {\bibinfo {volume} {21}},\ \bibinfo {pages}
  {1357--1376} (\bibinfo {year} {2007})}\BibitemShut {NoStop}%
\bibitem [{\citenamefont {Donk{\'{o}}}\ \emph {et~al.}(2009)\citenamefont
  {Donk{\'{o}}}, \citenamefont {Goree}, \citenamefont {Hartmann},\ and\
  \citenamefont {Liu}}]{DonkoPRE2009}%
  \BibitemOpen
  \bibfield  {author} {\bibinfo {author} {\bibfnamefont {Z.}~\bibnamefont
  {Donk{\'{o}}}}, \bibinfo {author} {\bibfnamefont {J.}~\bibnamefont {Goree}},
  \bibinfo {author} {\bibfnamefont {P.}~\bibnamefont {Hartmann}}, \ and\
  \bibinfo {author} {\bibfnamefont {B.}~\bibnamefont {Liu}},\ }\bibfield
  {title} {\enquote {\bibinfo {title} {Time-correlation functions and transport
  coefficients of two-dimensional yukawa liquids},}\ }\href {\doibase
  10.1103/physreve.79.026401} {\bibfield  {journal} {\bibinfo  {journal} {Phys.
  Rev. E}\ }\textbf {\bibinfo {volume} {79}},\ \bibinfo {pages} {026401}
  (\bibinfo {year} {2009})}\BibitemShut {NoStop}%
\bibitem [{\citenamefont {Lepri}\ \emph {et~al.}(2003)\citenamefont {Lepri},
  \citenamefont {Livi},\ and\ \citenamefont {Politi}}]{LepriPR2003}%
  \BibitemOpen
  \bibfield  {author} {\bibinfo {author} {\bibfnamefont {S.}~\bibnamefont
  {Lepri}}, \bibinfo {author} {\bibfnamefont {R.}~\bibnamefont {Livi}}, \ and\
  \bibinfo {author} {\bibfnamefont {A.}~\bibnamefont {Politi}},\ }\bibfield
  {title} {\enquote {\bibinfo {title} {Thermal conduction in classical
  low-dimensional lattices},}\ }\href {\doibase 10.1016/s0370-1573(02)00558-6}
  {\bibfield  {journal} {\bibinfo  {journal} {Phys. Rep.}\ }\textbf {\bibinfo
  {volume} {377}},\ \bibinfo {pages} {1--80} (\bibinfo {year}
  {2003})}\BibitemShut {NoStop}%
\bibitem [{\citenamefont {Nunomura}\ \emph {et~al.}(2005)\citenamefont
  {Nunomura}, \citenamefont {Samsonov}, \citenamefont {Zhdanov},\ and\
  \citenamefont {Morfill}}]{NunomuraPRL2005}%
  \BibitemOpen
  \bibfield  {author} {\bibinfo {author} {\bibfnamefont {S.}~\bibnamefont
  {Nunomura}}, \bibinfo {author} {\bibfnamefont {D.}~\bibnamefont {Samsonov}},
  \bibinfo {author} {\bibfnamefont {S.}~\bibnamefont {Zhdanov}}, \ and\
  \bibinfo {author} {\bibfnamefont {G.}~\bibnamefont {Morfill}},\ }\bibfield
  {title} {\enquote {\bibinfo {title} {Heat transfer in a two-dimensional
  crystalline complex (dusty) plasma},}\ }\href {\doibase
  10.1103/physrevlett.95.025003} {\bibfield  {journal} {\bibinfo  {journal}
  {Phys. Rev. Lett.}\ }\textbf {\bibinfo {volume} {95}},\ \bibinfo {pages}
  {025003} (\bibinfo {year} {2005})}\BibitemShut {NoStop}%
\bibitem [{\citenamefont {Fortov}\ \emph {et~al.}(2007)\citenamefont {Fortov},
  \citenamefont {Vaulina}, \citenamefont {Petrov}, \citenamefont {Vasiliev},
  \citenamefont {Gavrikov}, \citenamefont {Shakova}, \citenamefont {Vorona},
  \citenamefont {Khrustalyov}, \citenamefont {Manohin},\ and\ \citenamefont
  {Chernyshev}}]{FortovPRE2007}%
  \BibitemOpen
  \bibfield  {author} {\bibinfo {author} {\bibfnamefont {V.~E.}\ \bibnamefont
  {Fortov}}, \bibinfo {author} {\bibfnamefont {O.~S.}\ \bibnamefont {Vaulina}},
  \bibinfo {author} {\bibfnamefont {O.~F.}\ \bibnamefont {Petrov}}, \bibinfo
  {author} {\bibfnamefont {M.~N.}\ \bibnamefont {Vasiliev}}, \bibinfo {author}
  {\bibfnamefont {A.~V.}\ \bibnamefont {Gavrikov}}, \bibinfo {author}
  {\bibfnamefont {I.~A.}\ \bibnamefont {Shakova}}, \bibinfo {author}
  {\bibfnamefont {N.~A.}\ \bibnamefont {Vorona}}, \bibinfo {author}
  {\bibfnamefont {Yu.~V.}\ \bibnamefont {Khrustalyov}}, \bibinfo {author}
  {\bibfnamefont {A.~A.}\ \bibnamefont {Manohin}}, \ and\ \bibinfo {author}
  {\bibfnamefont {A.~V.}\ \bibnamefont {Chernyshev}},\ }\bibfield  {title}
  {\enquote {\bibinfo {title} {Experimental study of the heat transport
  processes in dusty plasma fluid},}\ }\href {\doibase
  10.1103/physreve.75.026403} {\bibfield  {journal} {\bibinfo  {journal} {Phys.
  Rev. E}\ }\textbf {\bibinfo {volume} {75}},\ \bibinfo {pages} {026403}
  (\bibinfo {year} {2007})}\BibitemShut {NoStop}%
\bibitem [{\citenamefont {Nosenko}\ \emph {et~al.}(2008)\citenamefont
  {Nosenko}, \citenamefont {Zhdanov}, \citenamefont {Ivlev}, \citenamefont
  {Morfill}, \citenamefont {Goree},\ and\ \citenamefont
  {Piel}}]{NosenkoPRL2008}%
  \BibitemOpen
  \bibfield  {author} {\bibinfo {author} {\bibfnamefont {V.}~\bibnamefont
  {Nosenko}}, \bibinfo {author} {\bibfnamefont {S.}~\bibnamefont {Zhdanov}},
  \bibinfo {author} {\bibfnamefont {A.~V.}\ \bibnamefont {Ivlev}}, \bibinfo
  {author} {\bibfnamefont {G.}~\bibnamefont {Morfill}}, \bibinfo {author}
  {\bibfnamefont {J.}~\bibnamefont {Goree}}, \ and\ \bibinfo {author}
  {\bibfnamefont {A.}~\bibnamefont {Piel}},\ }\bibfield  {title} {\enquote
  {\bibinfo {title} {Heat transport in a two-dimensional complex (dusty) plasma
  at melting conditions},}\ }\href {\doibase 10.1103/physrevlett.100.025003}
  {\bibfield  {journal} {\bibinfo  {journal} {Phys. Rev. Lett.}\ }\textbf
  {\bibinfo {volume} {100}},\ \bibinfo {pages} {025003} (\bibinfo {year}
  {2008})}\BibitemShut {NoStop}%
\bibitem [{\citenamefont {Goree}\ \emph {et~al.}(2013)\citenamefont {Goree},
  \citenamefont {Liu},\ and\ \citenamefont {Feng}}]{GoreePPCF2013}%
  \BibitemOpen
  \bibfield  {author} {\bibinfo {author} {\bibfnamefont {J.}~\bibnamefont
  {Goree}}, \bibinfo {author} {\bibfnamefont {B.}~\bibnamefont {Liu}}, \ and\
  \bibinfo {author} {\bibfnamefont {Y.}~\bibnamefont {Feng}},\ }\bibfield
  {title} {\enquote {\bibinfo {title} {Diagnostics for transport phenomena in
  strongly coupled dusty plasmas},}\ }\href {\doibase
  10.1088/0741-3335/55/12/124004} {\bibfield  {journal} {\bibinfo  {journal}
  {Plasma Phys. Control. Fusion}\ }\textbf {\bibinfo {volume} {55}},\ \bibinfo
  {pages} {124004} (\bibinfo {year} {2013})}\BibitemShut {NoStop}%
\bibitem [{\citenamefont {Du}\ \emph {et~al.}(2014)\citenamefont {Du},
  \citenamefont {Nosenko}, \citenamefont {Zhdanov}, \citenamefont {Thomas},\
  and\ \citenamefont {Morfill}}]{DuPRE2014}%
  \BibitemOpen
  \bibfield  {author} {\bibinfo {author} {\bibfnamefont {C.-R.}\ \bibnamefont
  {Du}}, \bibinfo {author} {\bibfnamefont {V.}~\bibnamefont {Nosenko}},
  \bibinfo {author} {\bibfnamefont {S.}~\bibnamefont {Zhdanov}}, \bibinfo
  {author} {\bibfnamefont {H.~M.}\ \bibnamefont {Thomas}}, \ and\ \bibinfo
  {author} {\bibfnamefont {G.~E.}\ \bibnamefont {Morfill}},\ }\bibfield
  {title} {\enquote {\bibinfo {title} {Channeling of particles and associated
  anomalous transport in a two-dimensional complex plasma crystal},}\ }\href
  {\doibase 10.1103/physreve.89.021101} {\bibfield  {journal} {\bibinfo
  {journal} {Phys. Rev. E}\ }\textbf {\bibinfo {volume} {89}},\ \bibinfo
  {pages} {021101} (\bibinfo {year} {2014})}\BibitemShut {NoStop}%
\bibitem [{\citenamefont {Hou}\ and\ \citenamefont {Piel}(2009)}]{HouJPA2009}%
  \BibitemOpen
  \bibfield  {author} {\bibinfo {author} {\bibfnamefont {L.-J.}\ \bibnamefont
  {Hou}}\ and\ \bibinfo {author} {\bibfnamefont {A.}~\bibnamefont {Piel}},\
  }\bibfield  {title} {\enquote {\bibinfo {title} {Heat conduction in 2d
  strongly coupled dusty plasmas},}\ }\href {\doibase
  10.1088/1751-8113/42/21/214025} {\bibfield  {journal} {\bibinfo  {journal}
  {J. Phys. A}\ }\textbf {\bibinfo {volume} {42}},\ \bibinfo {pages} {214025}
  (\bibinfo {year} {2009})}\BibitemShut {NoStop}%
\bibitem [{\citenamefont {Khrustalyov}\ and\ \citenamefont
  {Vaulina}(2012)}]{KhrustalyovPRE2012}%
  \BibitemOpen
  \bibfield  {author} {\bibinfo {author} {\bibfnamefont {Yu.~V.}\ \bibnamefont
  {Khrustalyov}}\ and\ \bibinfo {author} {\bibfnamefont {O.~S.}\ \bibnamefont
  {Vaulina}},\ }\bibfield  {title} {\enquote {\bibinfo {title} {Numerical
  simulations of thermal conductivity in dissipative two-dimensional yukawa
  systems},}\ }\href {\doibase 10.1103/physreve.85.046405} {\bibfield
  {journal} {\bibinfo  {journal} {Phys. Rev. E}\ }\textbf {\bibinfo {volume}
  {85}},\ \bibinfo {pages} {046405} (\bibinfo {year} {2012})}\BibitemShut
  {NoStop}%
\bibitem [{\citenamefont {Shahzad}\ and\ \citenamefont
  {He}(2015)}]{ShahzadPoP2015}%
  \BibitemOpen
  \bibfield  {author} {\bibinfo {author} {\bibfnamefont {A.}~\bibnamefont
  {Shahzad}}\ and\ \bibinfo {author} {\bibfnamefont {M.-G.}\ \bibnamefont
  {He}},\ }\bibfield  {title} {\enquote {\bibinfo {title} {Numerical experiment
  of thermal conductivity in two-dimensional yukawa liquids},}\ }\href
  {\doibase 10.1063/1.4938275} {\bibfield  {journal} {\bibinfo  {journal}
  {Phys. Plasmas}\ }\textbf {\bibinfo {volume} {22}},\ \bibinfo {pages}
  {123707} (\bibinfo {year} {2015})}\BibitemShut {NoStop}%
\bibitem [{\citenamefont {Horrocks}\ and\ \citenamefont
  {McLaughlin}(1960)}]{Horrocks1960}%
  \BibitemOpen
  \bibfield  {author} {\bibinfo {author} {\bibfnamefont {J.~K.}\ \bibnamefont
  {Horrocks}}\ and\ \bibinfo {author} {\bibfnamefont {E.}~\bibnamefont
  {McLaughlin}},\ }\bibfield  {title} {\enquote {\bibinfo {title} {Thermal
  conductivity of simple molecules in the condensed state},}\ }\href {\doibase
  10.1039/tf9605600206} {\bibfield  {journal} {\bibinfo  {journal} {Trans.
  Faraday Soc.}\ }\textbf {\bibinfo {volume} {56}},\ \bibinfo {pages} {206}
  (\bibinfo {year} {1960})}\BibitemShut {NoStop}%
\bibitem [{\citenamefont {Bridgman}(1923)}]{Bridgman1923}%
  \BibitemOpen
  \bibfield  {author} {\bibinfo {author} {\bibfnamefont {P.~W.}\ \bibnamefont
  {Bridgman}},\ }\bibfield  {title} {\enquote {\bibinfo {title} {The thermal
  conductivity of liquids under pressure},}\ }\href {\doibase 10.2307/20026073}
  {\bibfield  {journal} {\bibinfo  {journal} {PNAAS}\ }\textbf {\bibinfo
  {volume} {59}},\ \bibinfo {pages} {141} (\bibinfo {year} {1923})}\BibitemShut
  {NoStop}%
\bibitem [{\citenamefont {Bird}\ \emph {et~al.}(2002)\citenamefont {Bird},
  \citenamefont {Lightfoot},\ and\ \citenamefont {Stewart}}]{BirdBook}%
  \BibitemOpen
  \bibfield  {author} {\bibinfo {author} {\bibfnamefont {R.~B.}\ \bibnamefont
  {Bird}}, \bibinfo {author} {\bibfnamefont {E.~N.}\ \bibnamefont {Lightfoot}},
  \ and\ \bibinfo {author} {\bibfnamefont {W.~E.}\ \bibnamefont {Stewart}},\
  }\href@noop {} {\emph {\bibinfo {title} {Transport Phenomena -}}}\ (\bibinfo
  {publisher} {J. Wiley},\ \bibinfo {address} {New York},\ \bibinfo {year}
  {2002})\BibitemShut {NoStop}%
\bibitem [{\citenamefont {Murillo}(2000)}]{MurilloPRL2000}%
  \BibitemOpen
  \bibfield  {author} {\bibinfo {author} {\bibfnamefont {M.~S.}\ \bibnamefont
  {Murillo}},\ }\bibfield  {title} {\enquote {\bibinfo {title} {Critical wave
  vectors for transverse modes in strongly coupled dusty plasmas},}\ }\href
  {\doibase 10.1103/physrevlett.85.2514} {\bibfield  {journal} {\bibinfo
  {journal} {Phys. Rev. Lett.}\ }\textbf {\bibinfo {volume} {85}},\ \bibinfo
  {pages} {2514--2517} (\bibinfo {year} {2000})}\BibitemShut {NoStop}%
\bibitem [{\citenamefont {Goree}\ \emph {et~al.}(2012)\citenamefont {Goree},
  \citenamefont {Donk{\'{o}}},\ and\ \citenamefont {Hartmann}}]{GoreePRE2012}%
  \BibitemOpen
  \bibfield  {author} {\bibinfo {author} {\bibfnamefont {J.}~\bibnamefont
  {Goree}}, \bibinfo {author} {\bibfnamefont {Z.}~\bibnamefont {Donk{\'{o}}}},
  \ and\ \bibinfo {author} {\bibfnamefont {P.}~\bibnamefont {Hartmann}},\
  }\bibfield  {title} {\enquote {\bibinfo {title} {Cutoff wave number for shear
  waves and maxwell relaxation time in yukawa liquids},}\ }\href {\doibase
  10.1103/physreve.85.066401} {\bibfield  {journal} {\bibinfo  {journal} {Phys.
  Rev. E}\ }\textbf {\bibinfo {volume} {85}},\ \bibinfo {pages} {066401}
  (\bibinfo {year} {2012})}\BibitemShut {NoStop}%
\bibitem [{\citenamefont {Yang}\ \emph {et~al.}(2017)\citenamefont {Yang},
  \citenamefont {Dove}, \citenamefont {Brazhkin},\ and\ \citenamefont
  {Trachenko}}]{YangPRL2017}%
  \BibitemOpen
  \bibfield  {author} {\bibinfo {author} {\bibfnamefont {C.}~\bibnamefont
  {Yang}}, \bibinfo {author} {\bibfnamefont {M.{\hspace{0.167em}}T.}\
  \bibnamefont {Dove}}, \bibinfo {author} {\bibfnamefont
  {V.{\hspace{0.167em}}V.}\ \bibnamefont {Brazhkin}}, \ and\ \bibinfo {author}
  {\bibfnamefont {K.}~\bibnamefont {Trachenko}},\ }\bibfield  {title} {\enquote
  {\bibinfo {title} {Emergence and evolution of the k-gap in spectra of liquid
  and supercritical states},}\ }\href {\doibase 10.1103/physrevlett.118.215502}
  {\bibfield  {journal} {\bibinfo  {journal} {Phys. Rev. Lett.}\ }\textbf
  {\bibinfo {volume} {118}},\ \bibinfo {pages} {215502} (\bibinfo {year}
  {2017})}\BibitemShut {NoStop}%
\bibitem [{\citenamefont {Khrapak}\ \emph {et~al.}(2018)\citenamefont
  {Khrapak}, \citenamefont {Kryuchkov}, \citenamefont {Mistryukova},
  \citenamefont {Khrapak},\ and\ \citenamefont {Yurchenko}}]{KhrapakJCP2018}%
  \BibitemOpen
  \bibfield  {author} {\bibinfo {author} {\bibfnamefont {S.~A.}\ \bibnamefont
  {Khrapak}}, \bibinfo {author} {\bibfnamefont {N.~P.}\ \bibnamefont
  {Kryuchkov}}, \bibinfo {author} {\bibfnamefont {L.~A.}\ \bibnamefont
  {Mistryukova}}, \bibinfo {author} {\bibfnamefont {A.~G.}\ \bibnamefont
  {Khrapak}}, \ and\ \bibinfo {author} {\bibfnamefont {S.~O.}\ \bibnamefont
  {Yurchenko}},\ }\bibfield  {title} {\enquote {\bibinfo {title} {Collective
  modes of two-dimensional classical coulomb fluids},}\ }\href {\doibase
  10.1063/1.5050708} {\bibfield  {journal} {\bibinfo  {journal} {J. Chem.
  Phys.}\ }\textbf {\bibinfo {volume} {149}},\ \bibinfo {pages} {134114}
  (\bibinfo {year} {2018})}\BibitemShut {NoStop}%
\bibitem [{\citenamefont {Khrapak}\ \emph {et~al.}(2019)\citenamefont
  {Khrapak}, \citenamefont {Khrapak}, \citenamefont {Kryuchkov},\ and\
  \citenamefont {Yurchenko}}]{KhrapakJCP2019}%
  \BibitemOpen
  \bibfield  {author} {\bibinfo {author} {\bibfnamefont {S.~A.}\ \bibnamefont
  {Khrapak}}, \bibinfo {author} {\bibfnamefont {A.~G.}\ \bibnamefont
  {Khrapak}}, \bibinfo {author} {\bibfnamefont {N.~P.}\ \bibnamefont
  {Kryuchkov}}, \ and\ \bibinfo {author} {\bibfnamefont {S.~O.}\ \bibnamefont
  {Yurchenko}},\ }\bibfield  {title} {\enquote {\bibinfo {title} {Onset of
  transverse (shear) waves in strongly-coupled yukawa fluids},}\ }\href
  {\doibase 10.1063/1.5088141} {\bibfield  {journal} {\bibinfo  {journal} {J.
  Chem. Phys.}\ }\textbf {\bibinfo {volume} {150}},\ \bibinfo {pages} {104503}
  (\bibinfo {year} {2019})}\BibitemShut {NoStop}%
\bibitem [{\citenamefont {Kryuchkov}\ \emph {et~al.}(2019)\citenamefont
  {Kryuchkov}, \citenamefont {Mistryukova}, \citenamefont {Brazhkin},\ and\
  \citenamefont {Yurchenko}}]{KryuchkovSciRep2019}%
  \BibitemOpen
  \bibfield  {author} {\bibinfo {author} {\bibfnamefont {N.~P.}\ \bibnamefont
  {Kryuchkov}}, \bibinfo {author} {\bibfnamefont {L.~A.}\ \bibnamefont
  {Mistryukova}}, \bibinfo {author} {\bibfnamefont {V.~V.}\ \bibnamefont
  {Brazhkin}}, \ and\ \bibinfo {author} {\bibfnamefont {S.~O.}\ \bibnamefont
  {Yurchenko}},\ }\bibfield  {title} {\enquote {\bibinfo {title} {Excitation
  spectra in fluids: How to analyze them properly},}\ }\href {\doibase
  10.1038/s41598-019-46979-y} {\bibfield  {journal} {\bibinfo  {journal} {Sci.
  Rep.}\ }\textbf {\bibinfo {volume} {9}},\ \bibinfo {pages} {10483} (\bibinfo
  {year} {2019})}\BibitemShut {NoStop}%
\bibitem [{\citenamefont {Kryuchkov}\ \emph {et~al.}(2020)\citenamefont
  {Kryuchkov}, \citenamefont {Mistryukova}, \citenamefont {Sapelkin},
  \citenamefont {Brazhkin},\ and\ \citenamefont
  {Yurchenko}}]{KryuchkovPRL2020}%
  \BibitemOpen
  \bibfield  {author} {\bibinfo {author} {\bibfnamefont {N.~P.}\ \bibnamefont
  {Kryuchkov}}, \bibinfo {author} {\bibfnamefont {L.~A.}\ \bibnamefont
  {Mistryukova}}, \bibinfo {author} {\bibfnamefont {A.~V.}\ \bibnamefont
  {Sapelkin}}, \bibinfo {author} {\bibfnamefont {V.~V.}\ \bibnamefont
  {Brazhkin}}, \ and\ \bibinfo {author} {\bibfnamefont {S.~O.}\ \bibnamefont
  {Yurchenko}},\ }\bibfield  {title} {\enquote {\bibinfo {title} {Universal
  effect of excitation dispersion on the heat capacity and gapped states in
  fluids},}\ }\href {\doibase 10.1103/physrevlett.125.125501} {\bibfield
  {journal} {\bibinfo  {journal} {Phys. Rev. Lett.}\ }\textbf {\bibinfo
  {volume} {125}},\ \bibinfo {pages} {125501} (\bibinfo {year}
  {2020})}\BibitemShut {NoStop}%
\bibitem [{\citenamefont {Nosenko}(2020)}]{NosenkoPC}%
  \BibitemOpen
  \bibfield  {author} {\bibinfo {author} {\bibfnamefont {V.}~\bibnamefont
  {Nosenko}},\ }\href@noop {} {}\bibinfo {howpublished} {{Private
  Communication}} (\bibinfo {year} {2020})\BibitemShut {NoStop}%
\bibitem [{\citenamefont {Khrapak}\ and\ \citenamefont
  {Khrapak}(2016)}]{KhrapakCPP2016}%
  \BibitemOpen
  \bibfield  {author} {\bibinfo {author} {\bibfnamefont {S.~A.}\ \bibnamefont
  {Khrapak}}\ and\ \bibinfo {author} {\bibfnamefont {A.~G.}\ \bibnamefont
  {Khrapak}},\ }\bibfield  {title} {\enquote {\bibinfo {title} {Internal energy
  of the classical two- and three-dimensional one-component-plasma},}\ }\href
  {\doibase 10.1002/ctpp.201500104} {\bibfield  {journal} {\bibinfo  {journal}
  {Contrib. Plasma Phys.}\ }\textbf {\bibinfo {volume} {56}},\ \bibinfo {pages}
  {270--280} (\bibinfo {year} {2016})}\BibitemShut {NoStop}%
\bibitem [{\citenamefont {Hartmann}\ \emph {et~al.}(2005)\citenamefont
  {Hartmann}, \citenamefont {Kalman}, \citenamefont {Donk{\'{o}}},\ and\
  \citenamefont {Kutasi}}]{HartmannPRE2005}%
  \BibitemOpen
  \bibfield  {author} {\bibinfo {author} {\bibfnamefont {P.}~\bibnamefont
  {Hartmann}}, \bibinfo {author} {\bibfnamefont {G.~J.}\ \bibnamefont
  {Kalman}}, \bibinfo {author} {\bibfnamefont {Z.}~\bibnamefont {Donk{\'{o}}}},
  \ and\ \bibinfo {author} {\bibfnamefont {K.}~\bibnamefont {Kutasi}},\
  }\bibfield  {title} {\enquote {\bibinfo {title} {Equilibrium properties and
  phase diagram of two-dimensional yukawa systems},}\ }\href {\doibase
  10.1103/physreve.72.026409} {\bibfield  {journal} {\bibinfo  {journal} {Phys.
  Rev. E}\ }\textbf {\bibinfo {volume} {72}},\ \bibinfo {pages} {026409}
  (\bibinfo {year} {2005})}\BibitemShut {NoStop}%
\bibitem [{\citenamefont {Kryuchkov}\ \emph {et~al.}(2017)\citenamefont
  {Kryuchkov}, \citenamefont {Khrapak},\ and\ \citenamefont
  {Yurchenko}}]{KryuchkovJCP2017}%
  \BibitemOpen
  \bibfield  {author} {\bibinfo {author} {\bibfnamefont {N.~P.}\ \bibnamefont
  {Kryuchkov}}, \bibinfo {author} {\bibfnamefont {S.~A.}\ \bibnamefont
  {Khrapak}}, \ and\ \bibinfo {author} {\bibfnamefont {S.~O.}\ \bibnamefont
  {Yurchenko}},\ }\bibfield  {title} {\enquote {\bibinfo {title}
  {Thermodynamics of two-dimensional yukawa systems across coupling regimes},}\
  }\href {\doibase 10.1063/1.4979325} {\bibfield  {journal} {\bibinfo
  {journal} {J. Chem. Phys.}\ }\textbf {\bibinfo {volume} {146}},\ \bibinfo
  {pages} {134702} (\bibinfo {year} {2017})}\BibitemShut {NoStop}%
\bibitem [{\citenamefont {Khrapak}(2020)}]{KhrapakPRR2020}%
  \BibitemOpen
  \bibfield  {author} {\bibinfo {author} {\bibfnamefont {S.~A.}\ \bibnamefont
  {Khrapak}},\ }\bibfield  {title} {\enquote {\bibinfo {title} {Lindemann
  melting criterion in two dimensions},}\ }\href {\doibase
  10.1103/physrevresearch.2.012040} {\bibfield  {journal} {\bibinfo  {journal}
  {Phys. Rev. Research}\ }\textbf {\bibinfo {volume} {2}},\ \bibinfo {pages}
  {012040} (\bibinfo {year} {2020})}\BibitemShut {NoStop}%
\bibitem [{\citenamefont {Khrapak}\ and\ \citenamefont
  {Klumov}(2018)}]{KhrapakPoP2018}%
  \BibitemOpen
  \bibfield  {author} {\bibinfo {author} {\bibfnamefont {S.}~\bibnamefont
  {Khrapak}}\ and\ \bibinfo {author} {\bibfnamefont {B.}~\bibnamefont
  {Klumov}},\ }\bibfield  {title} {\enquote {\bibinfo {title} {High-frequency
  elastic moduli of two-dimensional yukawa fluids and solids},}\ }\href
  {\doibase 10.1063/1.5025396} {\bibfield  {journal} {\bibinfo  {journal}
  {Phys. Plasmas}\ }\textbf {\bibinfo {volume} {25}},\ \bibinfo {pages}
  {033706} (\bibinfo {year} {2018})}\BibitemShut {NoStop}%
\bibitem [{\citenamefont {Donko}\ \emph {et~al.}(2008)\citenamefont {Donko},
  \citenamefont {Kalman},\ and\ \citenamefont {Hartmann}}]{DonkoJPCM2008}%
  \BibitemOpen
  \bibfield  {author} {\bibinfo {author} {\bibfnamefont {Z.}~\bibnamefont
  {Donko}}, \bibinfo {author} {\bibfnamefont {G.~J.}\ \bibnamefont {Kalman}}, \
  and\ \bibinfo {author} {\bibfnamefont {P.}~\bibnamefont {Hartmann}},\
  }\bibfield  {title} {\enquote {\bibinfo {title} {Dynamical correlations and
  collective excitations of yukawa liquids},}\ }\href {\doibase
  10.1088/0953-8984/20/41/413101} {\bibfield  {journal} {\bibinfo  {journal}
  {J. Phys.: Condens. Matter}\ }\textbf {\bibinfo {volume} {20}},\ \bibinfo
  {pages} {413101} (\bibinfo {year} {2008})}\BibitemShut {NoStop}%
\bibitem [{\citenamefont {Khrapak}(2016)}]{KhrapakPoP2016_Relations}%
  \BibitemOpen
  \bibfield  {author} {\bibinfo {author} {\bibfnamefont {S.~A.}\ \bibnamefont
  {Khrapak}},\ }\bibfield  {title} {\enquote {\bibinfo {title} {Relations
  between the longitudinal and transverse sound velocities in strongly coupled
  yukawa fluids},}\ }\href {\doibase 10.1063/1.4942171} {\bibfield  {journal}
  {\bibinfo  {journal} {Phys. Plasmas}\ }\textbf {\bibinfo {volume} {23}},\
  \bibinfo {pages} {024504} (\bibinfo {year} {2016})}\BibitemShut {NoStop}%
\bibitem [{Gam()}]{Gamma}%
  \BibitemOpen
  \href@noop {} {}\bibinfo {note} {In Ref.~\cite{KhrustalyovPRE2012} the
  effective coupling parameter
  $\Gamma^*=1.5(Q^2/Tr_p)(1+\kappa+\kappa^2/2)e^{-\kappa}$ is used, which
  reaches the value $\Gamma_c^*\simeq 150$ at crystallization. This should be
  compared with $\Gamma = Q^2/Ta$ and $\Gamma_c\simeq 140$ at crystallization
  in the weakly screening limit ($\kappa\ll
  1$)~\cite{HartmannPRE2005,KhrapakCPP2016,KhrapakPRR2020}. Such a comparison
  yields $r_p/a\simeq 1.4$.}\BibitemShut {Stop}%
\bibitem [{\citenamefont {Salin}\ and\ \citenamefont
  {Caillol}(2003)}]{SalinPoP2003}%
  \BibitemOpen
  \bibfield  {author} {\bibinfo {author} {\bibfnamefont {G.}~\bibnamefont
  {Salin}}\ and\ \bibinfo {author} {\bibfnamefont {J.-M.}\ \bibnamefont
  {Caillol}},\ }\bibfield  {title} {\enquote {\bibinfo {title} {Equilibrium
  molecular dynamics simulations of the transport coefficients of the yukawa
  one component plasma},}\ }\href {\doibase 10.1063/1.1566749} {\bibfield
  {journal} {\bibinfo  {journal} {Phys. Plasmas}\ }\textbf {\bibinfo {volume}
  {10}},\ \bibinfo {pages} {1220--1230} (\bibinfo {year} {2003})}\BibitemShut
  {NoStop}%
\bibitem [{\citenamefont {Salin}\ and\ \citenamefont
  {Caillol}(2002)}]{SalinPRL2002}%
  \BibitemOpen
  \bibfield  {author} {\bibinfo {author} {\bibfnamefont {G.}~\bibnamefont
  {Salin}}\ and\ \bibinfo {author} {\bibfnamefont {J.-M.}\ \bibnamefont
  {Caillol}},\ }\bibfield  {title} {\enquote {\bibinfo {title} {Transport
  coefficients of the yukawa one-component plasma},}\ }\href {\doibase
  10.1103/physrevlett.88.065002} {\bibfield  {journal} {\bibinfo  {journal}
  {Phys. Rev. Lett.}\ }\textbf {\bibinfo {volume} {88}},\ \bibinfo {pages}
  {065002} (\bibinfo {year} {2002})}\BibitemShut {NoStop}%
\bibitem [{\citenamefont {Donk{\'{o}}}\ and\ \citenamefont
  {Hartmann}(2004)}]{DonkoPRE2004}%
  \BibitemOpen
  \bibfield  {author} {\bibinfo {author} {\bibfnamefont {Z.}~\bibnamefont
  {Donk{\'{o}}}}\ and\ \bibinfo {author} {\bibfnamefont {P.}~\bibnamefont
  {Hartmann}},\ }\bibfield  {title} {\enquote {\bibinfo {title} {Thermal
  conductivity of strongly coupled yukawa liquids},}\ }\href {\doibase
  10.1103/physreve.69.016405} {\bibfield  {journal} {\bibinfo  {journal} {Phys.
  Rev. E}\ }\textbf {\bibinfo {volume} {69}},\ \bibinfo {pages} {016405}
  (\bibinfo {year} {2004})}\BibitemShut {NoStop}%
\bibitem [{\citenamefont {Faussurier}\ and\ \citenamefont
  {Murillo}(2003)}]{FaussurierPRE2003}%
  \BibitemOpen
  \bibfield  {author} {\bibinfo {author} {\bibfnamefont {G.}~\bibnamefont
  {Faussurier}}\ and\ \bibinfo {author} {\bibfnamefont {M.~S.}\ \bibnamefont
  {Murillo}},\ }\bibfield  {title} {\enquote {\bibinfo {title}
  {Gibbs-bogolyubov inequality and transport properties for strongly coupled
  yukawa fluids},}\ }\href {\doibase 10.1103/physreve.67.046404} {\bibfield
  {journal} {\bibinfo  {journal} {Phys. Rev. E}\ }\textbf {\bibinfo {volume}
  {67}},\ \bibinfo {pages} {046404} (\bibinfo {year} {2003})}\BibitemShut
  {NoStop}%
\bibitem [{\citenamefont {Faussurier}(2004)}]{FaussurierPRE2004}%
  \BibitemOpen
  \bibfield  {author} {\bibinfo {author} {\bibfnamefont {G.}~\bibnamefont
  {Faussurier}},\ }\bibfield  {title} {\enquote {\bibinfo {title} {Description
  of strongly coupled yukawa fluids using the variational modified hypernetted
  chain approach},}\ }\href {\doibase 10.1103/physreve.69.066402} {\bibfield
  {journal} {\bibinfo  {journal} {Phys. Rev. E}\ }\textbf {\bibinfo {volume}
  {69}},\ \bibinfo {pages} {066402} (\bibinfo {year} {2004})}\BibitemShut
  {NoStop}%
\bibitem [{\citenamefont {Shahzad}\ and\ \citenamefont
  {He}(2012{\natexlab{a}})}]{ShahzadPoP2012}%
  \BibitemOpen
  \bibfield  {author} {\bibinfo {author} {\bibfnamefont {A.}~\bibnamefont
  {Shahzad}}\ and\ \bibinfo {author} {\bibfnamefont {M.-G.}\ \bibnamefont
  {He}},\ }\bibfield  {title} {\enquote {\bibinfo {title} {Thermal conductivity
  calculation of complex (dusty) plasmas},}\ }\href {\doibase
  10.1063/1.4748526} {\bibfield  {journal} {\bibinfo  {journal} {Phys.
  Plasmas}\ }\textbf {\bibinfo {volume} {19}},\ \bibinfo {pages} {083707}
  (\bibinfo {year} {2012}{\natexlab{a}})}\BibitemShut {NoStop}%
\bibitem [{\citenamefont {Shahzad}\ and\ \citenamefont
  {He}(2012{\natexlab{b}})}]{ShahzadCPP2012}%
  \BibitemOpen
  \bibfield  {author} {\bibinfo {author} {\bibfnamefont {A.}~\bibnamefont
  {Shahzad}}\ and\ \bibinfo {author} {\bibfnamefont {M.-G.}\ \bibnamefont
  {He}},\ }\bibfield  {title} {\enquote {\bibinfo {title} {Thermal conductivity
  of three-dimensional yukawa liquids (dusty plasmas)},}\ }\href {\doibase
  10.1002/ctpp.201200002} {\bibfield  {journal} {\bibinfo  {journal}
  {Contrib.Plasma Phys.}\ }\textbf {\bibinfo {volume} {52}},\ \bibinfo {pages}
  {667--675} (\bibinfo {year} {2012}{\natexlab{b}})}\BibitemShut {NoStop}%
\bibitem [{\citenamefont {Scheiner}\ and\ \citenamefont
  {Baalrud}(2019)}]{ScheinerPRE2019}%
  \BibitemOpen
  \bibfield  {author} {\bibinfo {author} {\bibfnamefont {B.}~\bibnamefont
  {Scheiner}}\ and\ \bibinfo {author} {\bibfnamefont {S.~D.}\ \bibnamefont
  {Baalrud}},\ }\bibfield  {title} {\enquote {\bibinfo {title} {Testing thermal
  conductivity models with equilibrium molecular dynamics simulations of the
  one-component plasma},}\ }\href {\doibase 10.1103/physreve.100.043206}
  {\bibfield  {journal} {\bibinfo  {journal} {Phys. Rev. E}\ }\textbf {\bibinfo
  {volume} {100}},\ \bibinfo {pages} {043206} (\bibinfo {year}
  {2019})}\BibitemShut {NoStop}%
\bibitem [{\citenamefont {Nosenko}\ \emph {et~al.}(2018)\citenamefont
  {Nosenko}, \citenamefont {Meyer}, \citenamefont {Zhdanov},\ and\
  \citenamefont {Thomas}}]{NosenkoAIP2018}%
  \BibitemOpen
  \bibfield  {author} {\bibinfo {author} {\bibfnamefont {V.}~\bibnamefont
  {Nosenko}}, \bibinfo {author} {\bibfnamefont {J.}~\bibnamefont {Meyer}},
  \bibinfo {author} {\bibfnamefont {S.~K.}\ \bibnamefont {Zhdanov}}, \ and\
  \bibinfo {author} {\bibfnamefont {H.~M.}\ \bibnamefont {Thomas}},\ }\bibfield
   {title} {\enquote {\bibinfo {title} {New radio-frequency setup for studying
  large 2d complex plasma crystals},}\ }\href {\doibase 10.1063/1.5064457}
  {\bibfield  {journal} {\bibinfo  {journal} {{AIP} Adv.}\ }\textbf {\bibinfo
  {volume} {8}},\ \bibinfo {pages} {125303} (\bibinfo {year}
  {2018})}\BibitemShut {NoStop}%
\bibitem [{\citenamefont {Alastuey}\ and\ \citenamefont
  {Jancovici}(1981)}]{Alastuey1981}%
  \BibitemOpen
  \bibfield  {author} {\bibinfo {author} {\bibfnamefont {A.}~\bibnamefont
  {Alastuey}}\ and\ \bibinfo {author} {\bibfnamefont {B.}~\bibnamefont
  {Jancovici}},\ }\bibfield  {title} {\enquote {\bibinfo {title} {On the
  classical two-dimensional one-component coulomb plasma},}\ }\href {\doibase
  10.1051/jphys:019810042010100} {\bibfield  {journal} {\bibinfo  {journal} {J.
  Phys.}\ }\textbf {\bibinfo {volume} {42}},\ \bibinfo {pages} {1--12}
  (\bibinfo {year} {1981})}\BibitemShut {NoStop}%
\bibitem [{\citenamefont {Caillol}\ \emph {et~al.}(1982)\citenamefont
  {Caillol}, \citenamefont {Levesque}, \citenamefont {Weis},\ and\
  \citenamefont {Hansen}}]{Caillol1982}%
  \BibitemOpen
  \bibfield  {author} {\bibinfo {author} {\bibfnamefont {J.~M.}\ \bibnamefont
  {Caillol}}, \bibinfo {author} {\bibfnamefont {D.}~\bibnamefont {Levesque}},
  \bibinfo {author} {\bibfnamefont {J.~J.}\ \bibnamefont {Weis}}, \ and\
  \bibinfo {author} {\bibfnamefont {J.~P.}\ \bibnamefont {Hansen}},\ }\bibfield
   {title} {\enquote {\bibinfo {title} {A monte carlo study of the classical
  two-dimensional one-component plasma},}\ }\href {\doibase 10.1007/bf01012609}
  {\bibfield  {journal} {\bibinfo  {journal} {J. Stat. Phys.}\ }\textbf
  {\bibinfo {volume} {28}},\ \bibinfo {pages} {325--349} (\bibinfo {year}
  {1982})}\BibitemShut {NoStop}%
\bibitem [{\citenamefont {de~Leeuw}\ and\ \citenamefont
  {Perram}(1982)}]{Leeuw1982}%
  \BibitemOpen
  \bibfield  {author} {\bibinfo {author} {\bibfnamefont {S.W.}\ \bibnamefont
  {de~Leeuw}}\ and\ \bibinfo {author} {\bibfnamefont {J.W.}\ \bibnamefont
  {Perram}},\ }\bibfield  {title} {\enquote {\bibinfo {title} {Statistical
  mechanics of two-dimensional coulomb systems: Ii. the two-dimensional
  one-component plasma},}\ }\href {\doibase 10.1016/0378-4371(82)90156-x}
  {\bibfield  {journal} {\bibinfo  {journal} {Phys. A}\ }\textbf {\bibinfo
  {volume} {113}},\ \bibinfo {pages} {546--558} (\bibinfo {year}
  {1982})}\BibitemShut {NoStop}%
\bibitem [{\citenamefont {de~Leeuw}\ \emph {et~al.}(1983)\citenamefont
  {de~Leeuw}, \citenamefont {Perram},\ and\ \citenamefont {Smith}}]{Leeuw1983}%
  \BibitemOpen
  \bibfield  {author} {\bibinfo {author} {\bibfnamefont {S.W.}\ \bibnamefont
  {de~Leeuw}}, \bibinfo {author} {\bibfnamefont {J.W.}\ \bibnamefont {Perram}},
  \ and\ \bibinfo {author} {\bibfnamefont {E.R.}\ \bibnamefont {Smith}},\
  }\bibfield  {title} {\enquote {\bibinfo {title} {Statistical mechanics of
  two-dimensional coulomb systems: Iii. dynamic properties of the
  two-dimensional one-component plasma},}\ }\href {\doibase
  10.1016/0378-4371(83)90102-4} {\bibfield  {journal} {\bibinfo  {journal}
  {Phys. A}\ }\textbf {\bibinfo {volume} {119}},\ \bibinfo {pages} {441--454}
  (\bibinfo {year} {1983})}\BibitemShut {NoStop}%
\bibitem [{\citenamefont {Khrapak}\ \emph {et~al.}(2016)\citenamefont
  {Khrapak}, \citenamefont {Klumov},\ and\ \citenamefont
  {Khrapak}}]{KhrapakPoP2016_2DOCP}%
  \BibitemOpen
  \bibfield  {author} {\bibinfo {author} {\bibfnamefont {S.~A.}\ \bibnamefont
  {Khrapak}}, \bibinfo {author} {\bibfnamefont {B.~A.}\ \bibnamefont {Klumov}},
  \ and\ \bibinfo {author} {\bibfnamefont {A.~G.}\ \bibnamefont {Khrapak}},\
  }\bibfield  {title} {\enquote {\bibinfo {title} {Collective modes in
  two-dimensional one-component-plasma with logarithmic interaction},}\ }\href
  {\doibase 10.1063/1.4950829} {\bibfield  {journal} {\bibinfo  {journal}
  {Phys. Plasmas}\ }\textbf {\bibinfo {volume} {23}},\ \bibinfo {pages}
  {052115} (\bibinfo {year} {2016})}\BibitemShut {NoStop}%
\end{thebibliography}%

\end{document}